\documentstyle[12pt,a4]{article}
\author{Yu.~V.~Lipko, A.~Yu.~Pasinin, R.~A.~Rakhmatulin\\
        Institute of Solar-Terrestrial Physics SD RAS,\\
        p.~o.~box~4026, Irkutsk, 664033, Russia\\
        fax: +7 3952 462557; e-mail:~lipko@iszf.irk.ru}

\title{IONOSPHERIC MANIFESTATIONS OF GEOMAGNETIC PULSATIONS
       AT HIGH LATITUDES}
\date{}
\begin{document}
\sloppy
\maketitle
\begin{abstract}
In this paper the interrelation between geomagnetic pulsations and
variations in frequency Doppler shift $f_d$ of the ionosphere-reflected
radio signal is under investigation. The experiment on simultaneous
recording of $f_d$ variations and geomagnetic pulsations was organised at
high latitude station in Norilsk (geomagnetic latitude and longitude
$64.2^\circ$N, $160.4^\circ$E, $L=5.3$) during Febrary-April of 1995-98. Thirty
cases of simultaneous recording of duration from 10 min to two hour
were analysed: 6 cases of simultaneous recording of variations $f_d$ and
regular geomagnetic pulsations Pc5; and 25 cases of recording of $f_d$
variations and irregular pulsations Pi2.

On the basis of experimental results, the following conclusions have
been drawn: a) Hydromagnetic waves in the range of regular Pc5
pulsations, when interacting with the ionospheric F2 layer, make the
main contribution to short-period $f_d$ variations. The possible
mechanism of $f_d$ variations are oscillations of electron density,
associated with distribution of a hydromagnetic wave in an ionosphere.
b) There exists an unquestionable interrelation between $f_d$ variations
of the sporadic E layer-reflected radio signal and irregular Pi2
pulsations, but for some reasons it is traced poorly. 
\end{abstract}

\section{Introduction}
\label{intr}
The correlation between variations of ionospheric parameters and
variations of the Earth's geomagnetic field has attracted the
attention of researchers over decades [1, 2, 3]. Comparison of
fragments of simultaneously recorded geomagnetic pulsations and
short-period variations in frequency Doppler shift $f_d$ of the
ionosphere-reflected radio signal point to an unquestionable
correlation between ionospheric parameter variations and the
excitation regime of geomagnetic pulsations. The possibility of
investigating Pi2 geomagnetic pulsations by recording $f_d$ variations is
discussed in [4, 5]. In [5], it is confirmed that geomagnetic
pulsations are hydromagnetic waves generated in the magnetosphere. The
authors of [5, 6] have established that hydromagnetic waves in the
ionosphere can be studied by using the Doppler method. In [7], it is
suggested that hydromagnetic waves give rise to a displacement of the
ionospheric layer, and hence to the appearance in the ionosphere of
small-scale wave disturbances with frequencies ranging from $1.5$ mHz to
100-200 mHz. It is also argued that there is a close relationship
between short-period disturbances in the ionosphere and Pi2, Pip and
PiC geomagnetic pulsations [7]. The authors of [8] have revealed that
wave disturbances in the high-latitude ionosphere are accompanied by
the generation of Pi2 pulsations.

In spite of many years of theoretical and experimental research, it
has not yet been possible to conclusively elucidate the mechanisms [6,
9, 10] accounting for the relationship between short-period variations
in frequency Doppler shift of the ionosphere-reflected radio signal
and geomagnetic pulsations. One of the reasons for this, as suggested
by the authors of [10], has to do with the difficulty in obtaining
statistically significant sets of experimental data.

Thus the
relationship between $f_d$ variations and geomagnetic pulsations is of
profound importance for the investigation of the influence of
hydromagnetic waves on the structure and dynamics of the high-latitude
ionosphere, on the one hand, and for the study of the geomagnetic
pulsations themselves, on the other. Unfortunately, currently there is
no clear understanding both of the morphology and of the physical
origin of the relationships between ionospheric parameter variations
and geomagnetic pulsations. The objective of this experimental paper
is to establish the behavior patterns of the high-latitude ionosphere
during periods of observation of different kinds of geomagnetic
pulsations.

During 1995-1998, at the high-latitude station in Norilsk
(geomagnetic latitude and longitude $64.2^\circ$N, $160.4^\circ$E,
$L=5.3$), an
experiment was organized on a simultaneous recording of $f_d$ variations
and geomagnetic pulsations.

\section{Experimental equipment and technique}
\label{Equipment}
The measurements of ionospheric parameters were made with the
hardware-software facility for vertical-incidence ionospheric sounding
that was developed on the basis of ionosonde R-017 and a personal
computer [11]. Some characteristics of the facility are: 2 kW
transmitter pulse power, $135\pm 15$ ms pulse duration, and 50 Hz sounding
pulse repetition rate.

The impulse signal in the vertical-incidence
sounding mode was transmitted using the broadband delta-type antenna,
with the main lobe width of the beam of 600. Half-wave dipole antennas
operating at the mean frequency of the working range, 6 MHz, were used
as receive antennas.

When operated at fixed frequency, the facility provided a Doppler
spectrum of the ionosphere-reflected radio signal, as well as making
it possible to calculate the weighted mean frequency Doppler shift $f_d$.
The interval for spectral analysis was taken to be 20 s, which
ensured a frequency resolution of $0.05$ Hz. A spectral analysis was
carried out in the frequency band $\pm 6$ Hz. Observations as long as 15
min to several hours were carried out with the signal/noise ratio of
at least 10. The experiment used working frequencies in the range 2-4
mHz; the signal reflected from the ionospheric $E_s$ and F2 layers was
received. Geomagnetic pulsations were recorded on a 24-hour basis with
the induction magnetometer in the frequency band 0.5-0.005 Hz, in the
dynamic range 0.01-100 nT.

\section{Experimental results}
\label{Experiment}

The observations were carried out in March, 1995, February, 1996, in
March-April 1998. The $f_d$ variations were recorded during passage
of the high latitude geomagnetic pulsations (GP): regular pulsations
Pc5 and irregular pulsations Pi2. In total was analysed about 30 cases
of simultaneous recording of $f_d$ variations and GP. From them six
cases were of recording $f_d$ variations during passage of pulsations
Pc5 and 25 cases were of recording of $f_d$ variations during passage
Pi2.

Let's consider in more detail two cases of simultaneous recording of
$f_d$ variations and regular pulsations Pc5 which were observed in
April 1, 1998 at 12{:}38-13{:}00 UT (18{:}38-19{:}00 LT) and in April 5, 1998
at 11{:}20-11{:}40 UT (17{:}20-17{:}40 LT). The geomagnetic conditions of
April 1, 1998 were quiet, the coefficient of geomagnetic activity Kp
was equal 1. The amplitude north-south component Pc5 was 33 nT, period
of oscillations -- 350 s (Fig{.} 1 a). Amplitude by east-west component
Pc5 -- 30 nT, principal periods of oscillations -- 210 s and 480 s (Fig{.}
1 b). The sounding was conducted on frequency $4.5$ MHz at a relation
signal/noise not less than 10. The signal reflected from a F2-layer of
ionosphere was accepted. The critical frequency $f_0F2$ was 6-7 MHz,
effective height $h'$ -- 270-300 km. Amplitude $f_d$ was $0.49$ Hz, the
periods -- 210 s and 480 s. Fig{.} 1 d, e, f submit the diagrams of the
spectrum time analysis (STAN) at coordinates time-period-spectral
density (d) -- north-south (N-S) component of GP; (e) -- east-west (E-W)
component of GP; (f) -- variations $f_d$. The variations of $f_d$ respond to
occurrence of geomagnetic pulsations and they are in an antiphase with
oscillations of east-west component GP. Thus the variations $f_d$ react
not only to pulsation of major amplitude (30 nT at 12{:}38 UT), but also
ones respond to on feeble oscillations 7 nT at 12{:}10 UT. The diagrams
STAN submitted in Fig{.} 1 d, e, f clearly show this fact. The response
of $f_d$ variations on feeble oscillations GP is imposed on $f_d$ variation
with period approximately 30 min. The $f_d$ variations with such periods
is explain by passage of acoustic-gravity waves (AGW), which periods
for heights of a F2-layer lies in the range between 10 and 40 min.

The regular pulsations Pc5 in April 5, 1998 also were observed under
quiet geomagnetic requirements (Kp=1) (Fig{.} 2). The amplitude of
north-south Pc5 component was $16.5$ nT, period of oscillations -- 360 s
(Fig{.} 2 a). Amplitude of east-west Pc5 component -- $9.5 nT$, period of
oscillations -- 360 s (Fig{.} 2 b). Variations of $f_d$ are submitted on
Fig{.} 2 c. The sounding operation was performed at 5 MHz frequency at a
relation signal/noise not less than 15-20. The radiosignal reflected
from a F2 layer of an ionosphere was accepted. The critical frequency
$f_0F2$ was 7-8 MHz, effective height $h'$ -- 270 km. The amplitude of $f_d$
variations was $0.56$ Hz, a period of oscillations -- 360 s. As well as
in case of April 5 the long-period variations with a period of
medium-scale TIDs ($1.7$ mHz) are superimposed on $f_d$ variations. It
clearly that $f_d$ variations correlate with oscillations of GP. The
variations $f_d$ are in a phase with oscillations of east-west GP
component and are displaced in phase by 2-3 minutes for the
north-south components of GP oscillations. The principal frequency of
$f_d$ variations coincides with frequency of both north-south and
east-west components of GP oscillations.

During observation of regular Pc5 pulsations in 5 of six cases the
good correlation of time series both spectrums of pulsations and $f_d$
variations is observed; for one or for both components of the field of
pulsations. In each individual case the responses of an ionosphere on
GP oscillations have some different. Specifically the $f_d$ variations
can be react on both components of GP or on one of them.

Irregularity pulsations Pi2 are observed in the explosive phase of
geomagnetic substorm. Norilsk at this time was in auroral zone. The
regular layer F2 was screened by a sporadic layer …. The severity of
the recording problem implied that a total absorption of radio waves
is generally observed in the ionosphere, when the strongest,
longest-lasting geomagnetic pulsations are excited. Nevertheless the
most part of observational data (25 cases) were received for
pulsations Pi2.

In six cases from these 25 the conformity between Pi2 and $f_d$ is
marked. Fig{.} 3 present the case in February 15, 1996 at 19{:}25-19{:}44
UT. At generation Pi2 there is a abrupt amplification of $f_d$
variations. The maximum amplitude of north-south Pi2 component reach
21 nT, principal periods of oscillations -- 50 s and 160 s (Fig{.} 3 a).
Amplitude of east-west Pi2 component was 15 nT, the principal periods
of oscillations -- 65 s and 120 s (Fig{.} 3 b). Fig{.} 3 c presents $f_d$
variations. The sounding was conducted on frequency 3 MHz at a
relation signal/noise about 5. The radiosignal reflected from
$E_{sr}$-layer was accepted. The amplitude of $f_d$ variations at the moment
of oscillation Pi2 sharply has increased to $0.34$ Hz. The principal
periods of $f_d$ variations were 45 s, 65 s, 120 s, 180 s. Results for
other five cases are simultaneous. At occurrence irregularity
pulsations Pi2 the amplitude of $f_d$ variations was incremented and in
the spectrum of variations $f_d$ there are spectral components with GP
periods. The spectrum of $f_d$ variations is the complex, it include more
spectral components, than spectrum GP. It is explained by that
variation of parameters of the ionosphere cause not only geomagnetic
pulsations, but also series of other factors.

In the majority of cases was revealed not any of conformity between $f_d$
variations and irregularity pulsations neither in a time series, nor
in spectrums. One of such cases (March 27, 1998) is given in Fig{.} 4.
In these cases on $f_d$ variations influence mechanisms more strong, than
mechanisms associated with GP.

\section{Discussion}
\label{Discussion}
It is universally accepted that geomagnetic pulsations, observed on
the ground, represent the manifestations of magnetospheric
hydrodynamic waves propagating downward through the ionosphere [11, 13].
It is common knowledge that the ionosphere modifies the
characteristics of hydromagnetic waves [13]. On the other hand, ionospheric
parameters can also undergo changes under the effect of hydrodynamic
waves. This is confirmed by simultaneous recordings of variations in
frequency Doppler shift $f_d$ of the ionosphere-reflected radio signal
and geomagnetic pulsations [1-3, 6, 9, 10, 14]. The time coincidence of $f_d$
variations and geomagnetic pulsations suggests that hydromagnetic
waves in the range of pulsations have a pronounced effect on the
ionosphere.

It should be noted that the mechanisms accounting for the correlation
between geomagnetic pulsations and $f_d$ variations still remain unclear
[9, 10].

As has been pointed out above, the time series of Pc5 pulsations and
$f_d$ variations have a high degree of correlation (Fig{.} 1, 2). The
ratio of amplitudes of $f_d$ variations and geomagnetic pulsations was
$0.02-0.07$ Hz/nT. This value is consistent with results (from $0.01$
Hz/nT to $0.4$ Hz/nT) obtained in [2, 10]. It is clearly seen from
Fig{.} 2b and 2c that at the time of observation of Pc5 pulsations the
spectral component of the $f_d$ variations, corresponding to the
frequency of Pc5, increases abruptly. The agreement between the time
series and spectra of the simultaneously recorded geomagnetic
pulsations and $f_d$ variations suggests that hydromagnetic waves of
the Pc5 range, when interacting with the ionospheric F2 layer, make
the main contribution to short-period $f_d$ variations.

As in [2, 3, 6], in most cases we detected only a slight correlation
between $f_d$ variations and geomagnetic pulsations. The simultaneous
recording of Pi2 geomagnetic pulsations and $f_d$ variations shows: a) at
all events an almost total absence of correlation between the time
series; and b) in series of cases coincidence of the principal maxima
on the plots of spectral density of these phenomena at the time of Pi2
observation. A similar situation was observed in the case of a
simultaneous recording of the luminosity of polar auroras and Pi2
pulsations [15]. The absence of correlation between time series of
irregular Pi2 pulsations and $f_d$ variations of the $E_s$ layer-reflected
radio signal may be attributed to a variety of factors. In the first
place, in high latitudes there may exist several sources of Pi2 spaced
by several degrees in latitude [15]. Secondly, at the sporadic E layer
heights the density of the neutral component is significantly higher,
and its influence on ionospheric plasma drifts is stronger when
compared with the F2 layer. And thirdly, magnetic variations may be
the response of ionospheric currents within the field of view of
magnetometers (within the range of several hundred kilometres),
whereas the ionosonde measures less distant disturbances in the
ionosphere over the observation site, within the range of 50-100 km
[4]. Furthermore, there may be interference of the reflected radio
waves from ionospheric irregularities, as pointed out in [3, 4].

Nevertheless, it's suggest that there is unquestionable correlation
between $f_d$ variations of the sporadic E layer-reflected radio signal
and irregular Pi2 pulsations.

\section{Conclusions}
\label{Conclusions}
An experiment has been conducted on a simultaneous recording of the
variations in frequency Doppler shift $f_d$ of the ionosphere-reflected
radio signal and geomagnetic pulsations measured on the ground.

Experimental results have revealed the following basic features:

a) the presence of a good correlation between regular Pc5 geomagnetic
pulsations and variations in frequency Doppler shift $f_d$ of the
ionospheric F2 layer-reflected radio signal;

b) the abrupt enhancement
of the spectral component of variations $f_d$ at the frequency coinciding
with that pulsations, at the time of Pc5 observation;

c) absence of a
correlation between irregular Pi2 geomagnetic pulsations and
variations in frequency Doppler shift of the radio signal reflected
from the sporadic E layer of the auroral ionosphere;

d) in the series
of cases at the time of Pi2 observation - an enhancement of the
short-period components of the spectrum of $f_d$ variations, and the
coincidence of the principal maxima on the plots of spectral density
of $f_d$ variations and Pi2.

On the basis of these main results, the
following conclusions have been drawn:

Hydromagnetic waves in the
range of regular Pc5 pulsations, when interacting with the ionospheric
F2 layer, make the main contribution to short-period $f_d$ variations.
The possible mechanism of $f_d$ variations are oscillations of electron
density, associated with distribution of a hydromagnetic wave in an
ionosphere.

There exists an unquestionable interrelation between $f_d$
variations of the sporadic E layer-reflected radio signal and
irregular Pi2 pulsations, but for some reasons it is noted poorly.

\section{Acknowledgements}
\label{Acnow}

We are indebted to V.V.Klimenko and to all staff members of the
Norilsk station for assistance in conducting the experiment. Special
thanks are due to E.L. Afraimovich and G.A. Zherebtsov for useful
discussions of the results.

\end{document}